\documentclass[conference]{IEEEtran}%
\usepackage{amsmath,amssymb,amsfonts,latexsym,multirow,graphicx}
\usepackage{epsfig}
\newcommand{\rk}[1]{\textup{\mbox{\,rank}}\,(#1)}
\newcommand{\Proof}{\noindent{\bf Proof.}\quad}
\newcommand{\qed}{\hfill$\Box$}
\newtheorem{theorem}{Theorem}[section]

\def\0{{\mathbf{0}}}
\def\1{{\mathbf{1}}}
\def\2{{\mathbf{2}}}
\def\3{{\mathbf{3}}}
\def\4{{\mathbf{4}}}
\def\5{{\mathbf{5}}}
\def\6{{\mathbf{6}}}
\def\7{{\mathbf{7}}}
\def\8{{\mathbf{8}}}
\def\9{{\mathbf{9}}}

\begin{document}


\title{Adaptively correcting quantum errors with entanglement}

\author{\IEEEauthorblockN{Yuichiro Fujiwara}
\IEEEauthorblockA{Department of Mathematical Sciences\\ Michigan Technological University\\ Houghton, MI 49931, USA\\
Email: yfujiwar@mtu.edu}
\and
\IEEEauthorblockN{Min-Hsiu Hsieh}
\IEEEauthorblockA{Statistical Laboratory\\ University of Cambridge\\ Wilberforce Road, Cambridge CB3 0WB, UK\\
Email: minhsiuh@gmail.com}}

\maketitle

\begin{abstract}
Contrary to the assumption that most quantum error-correcting codes (QECC) make,
it is expected that phase errors are much more likely than bit errors in physical devices.
By employing the entanglement-assisted stabilizer formalism,
we develop a new kind of error-correcting protocol which can flexibly trade
error correction abilities between the two types of errors, such that high error correction performance is achieved both in symmetric
and in asymmetric situations.
The characteristics of the QECCs can be optimized in an adaptive manner during information transmission.
The proposed entanglement-assisted QECCs require only one ebit regardless of the degree of asymmetry at a given moment
and can be decoded in polynomial time.
\end{abstract}

\section{Introduction}\label{sec:introduction}
The development of quantum error-correcting codes (QECCs) allows one to detect and correct quantum errors caused by the imperfection of physical devices \cite{CRSS,Gdissertation}.
However, most of the known QECCs assume that phase errors and bit errors occur equally likely. Recent investigation demonstrates that the noise in physical devices is typically asymmetric; phase errors are far more likely than bit errors in most situations (see, for example, \cite{IM} and references therein).

The effect can be catastrophic if a QECC designed for symmetric phase and bit errors is employed in a quantum system whose components produce asymmetric errors. The type of errors that happens more often will overwhelm the overall system performance while the error-correcting ability for the less likely type of errors will be wasted.
In fact, it is shown that QECCs taking advantage of the asymmetry in quantum errors achieve significantly better error correction performance \cite{IM,SEDH,SKR,WFLX}.
However, there has been made little progress on constructions for codes that can be adaptively fine-tuned according to the degree of asymmetry.

The primary purpose of this Letter is to develop an adaptive code construction, which can achieve very satisfactory performance and can flexibly trade error correction abilities between phase errors and bit errors. We employ a recently developed framework, called \textit{entanglement-assisted stabilizer formalism} \cite{BDH},
so that high error correction performance can be achieved in various situations ranging from symmetric to moderately asymmetric to heavily asymmetric.
The flexibility of error correction abilities is realized in such a way that one can optimize the characteristics of a QECC
in an adaptive manner during information transmission.

Our primary tools are the theories of low-density parity-check (LDPC) codes \cite{RUBook} and combinatorial designs \cite{BJLBook}.
Classical LDPC codes belong to an important class of modern coding theory. They can be systematically constructed using combinatorial design theory while almost achieving classical Shannon limit with very simple decoding circuits \cite{KLF}. Similar results have been recently observed in the quantum domain thanks to the ability of the entanglement-assisted stabilizer formalism to import every classical linear code \cite{BDH,DBH}. Combinatorial quantum LDPC codes based on the entanglement-assisted stabilizer formalism have the best error correction performance over the depolarizing channel \cite{HBD,HYH,FCVBT}. We will show that the adaptable noise control can be achieved by effectively utilizing LDPC codes and combinatorial mathematics. The proposed codes inherit the very low decoding complexity and notable high performance of entanglement-assisted quantum LDPC codes while requiring extremely small amounts of entanglement.

%

\section{Entanglement-assisted quantum LDPC codes}\label{sec:EA-LDPC}
QECCs based on the entanglement-assisted stabilizer formalism are called \textit{entanglement-assisted quantum error-correcting codes} (EAQECCs). An $[[n,k;c]]$ EAQECC encodes $k$ logical qubits into $n$ physical qubits with the help of $c$ copies of maximally entangled states ($c$ ebits). It has been shown that the great performance of EAQECCs does not necessarily come from the amount of ebits \cite{HYH,FCVBT}. Therefore, we will focus on EAQECCs requiring only one ebit in the Letter.

The CSS construction is one of the simplest methods to obtain quantum analogues of LDPC codes from binary linear codes \cite{DBH}.
The \textit{quantum check matrix} of a CSS-type EAQECC of length $n$ is of the form
\[\left[\begin{array}{cc}
H_1 & 0 \\
0 & H_2
\end{array}\right],\]
where $H_1$ and $H_2$ are parity-check matrices of binary linear codes of length $n$.
If $H_1$ and $H_2$ give $[n,k_1,d_1]$ and $[n,k_2,d_2]$ binary linear codes respectively,
then the resulting CSS-type EAQECC requires $c = \rk{H_1H_2^T}$ ebits and can encode $k_1 + k_2 - n + c$ logical qubits into $n$ physical qubits (See \cite{WB,HDB}).
A bounded-distance decoder can correct up to $\lfloor\frac{d_1-1}{2}\rfloor$ phase flips (Zs) and up to $\lfloor\frac{d_2-1}{2}\rfloor$ bit flips (Xs)
in two separate decoding steps.
We say that the code is of $Z$-\textit{distance} $d_1$ and $X$-\textit{distance} $d_2$.
We denote such a CSS-type EAQECC which is of length $n$, dimension $k$, $Z$-distance $d_z$, and $X$-distance $d_x$ and requires $c$ ebits
by $[[n,k,(d_z, d_x);c]]$.

The CSS construction often assumes that the parity-check matrices $H_1$ and $H_2$ are full-rank.
However, decoders of LDPC codes can exploit redundant rows in parity-check matrices without additional quantum interactions.
It is also readily checked that adding or deleting redundant rows does not change the required amount of entanglement.
For this reason, we allow linearly dependent rows in $H_1$ and $H_2$.
\begin{theorem}\label{th:CSS}
Let $H_1$ and $H_2$ be parity-check matrices of binary linear codes of parameters $[n,k_1,d_1]$ and $[n,k_2,d_2]$ respectively.
Then there exists an $[[n, k_1 + k_2 - n + c,(d_1,d_2);c]]$ \textup{EAQECC} with $c = \rk{H_1H_2^T}$.
\end{theorem}
If $H_1$ and $H_2$ have only small numbers of ones, the corresponding quantum check matrix can be efficiently decoded by the sum-product algorithm,
which qualifies the resulting EAQECC as an entanglement-assisted quantum LDPC code.
For details of entanglement-assisted quantum LDPC codes requiring only one ebit, the reader is referred to \cite{HYH,FCVBT} and references therein.

In the reminder of this section we define basic notions related to LDPC codes required in the subsequent sections.
For facts and undefined notions related to LDPC codes and combinatorial design theory, we refer the reader to \cite{RUBook,BJLBook}.
An LDPC code is \textit{regular} if its parity-check matrix has constant row and column weights.
Generally speaking, regular LDPC codes have better error floors than irregular ones.
One can optimize the threshold of an irregular LDPC code by a careful choice of row and column weights
in exchange for the performance in the error floor region.
In order to provide stable performance from adaptive noise control, we only employ regular LDPC codes.

A $4$-cycle in a parity-check matrix is a $2 \times 2$ all one sub-matrix. A $6$-cycle is a $3 \times 3$ sub-matrix in which each row and column has exactly two ones.
Typically short cycles negatively affect error correction performance.
If the shortest cycles in a parity-check matrix is of length $w$, the corresponding LDPC code is said to have \textit{girth} $w$.
The positive effect of avoiding $6$- or longer cycles is much smaller than avoiding $4$-cycles while it severely limits the available codes in the quantum setting.
For this reason, we focus on LDPC codes with girth six.

\section{Asymmetric EAQECCs}\label{sec:adaptive noise control}
In general, adding more rows to a parity-check matrix increases its error-correcting ability as long as the additional rows do not induce undesirable topological structures such as short cycles. We exploit this fact and the structure of the quantum check matrix
of a CSS-type EAQECC.

A pair of binary linear codes are \textit{isomorphic} if one can be obtained by permuting the coordinate positions of the other.
Let $H_1$ and $H_2$ be $v \times n$ matrices defining binary linear codes which are isomorphic but not identical and write
\[H_1 = \left[\begin{array}{c}
\boldsymbol{r}_1\\
\vdots\\
\boldsymbol{r}_v
\end{array}\right], H_2 = \left[\begin{array}{c}
\boldsymbol{s}_1\\
\vdots\\
\boldsymbol{s}_v
\end{array}\right],\]
where $\boldsymbol{r}_i$ and $\boldsymbol{s}_i$ are the $n$-dimensional binary vectors representing parity-check equations.
Because $H_1$ and $H_2$ define isomorphic codes,
the $Z$-distance and $X$-distance of the corresponding CSS-type code are the same.
If $H_1$ is obtained by permuting the rows and columns of $H_2$,
the sum-product algorithm gives the identical error correction performance for phase errors and bit errors.

The fundamental of our adaptive noise control is to swap part of rows responsible for phase errors and bit errors.
Construct a $w \times n$ binary matrix $R$ by taking $w$ rows of $H_2$.
Without loss of generality, we assume $R = \{\boldsymbol{s}_i | 1 \leq i \leq w\}$.
If discrepancy between assumed error probabilities and the actual channel behavior is detected,
we define a new quantum check matrix $H'$ as follows:
\[H' =  \left[\begin{array}{cc}
H_1'&0\\
0&H_2'
\end{array}\right]
=
 \left[\begin{array}{cc}
H_1&\multirow{2}{*}{0}\\
R&\\
\multirow{3}{*}{0}&\boldsymbol{s}_{w+1}\\
&\vdots\\
&\boldsymbol{s}_v
\end{array}\right].\]
$H'$ gives a CSS-type $[[n,n-\rk{H_1'}-\rk{H_2'}+\rk{H_1'H_2'^{T}};\rk{H_1'H_2'^{T}}]]$ EAQECC.

One may expect that in general the additional parity-check equations $R$ will increase the error correction ability for phase errors
as long as swapping rows does not induce $4$-cycles while bit errors will be less likely to be corrected because of the loss of the rows.
Ideally we would like $\rk{H_1} + \rk{H_2} = \rk{H_1'} + \rk{H_2'}$ to maintain the same dimension and to ensure the improvement
on the error correction ability for phase errors.
In order to keep the consumption of ebits as low as possible, $R$ must be chosen so that $\rk{H_1H_2^T} = \rk{H_1'H_2'^T} = 1$.


Thus, the criteria for desirable parity-check matrices $H_1$ and $H_2$ for adaptive noise control are summarized as follows:
\begin{itemize}
\item[1)]{$H_1$ and $H_2$ define isomorphic but not identical classical LDPC codes with girth six or greater,}
\item[2)]{\rk{$H_1H_2^T} = 1$,}
\item[3)]{$H_1$ and $H_2$ allow various choices of a set $R$ of rows
such that removing $R$ from $H_2$ and adding it to $H_1$ give $\rk{H_1'H_2'^T} = 1$ without inducing $4$-cycles,}
\item[4)]{$\rk{H_1} + \rk{H_2} = \rk{H_1'} + \rk{H_2'}$ for various choices of $R$ meeting Criterion 3).}
\end{itemize}
If the code designer wishes to utilize regular LDPC codes for each level of asymmetry, the pair of matrices must also satisfy:
\begin{itemize}
\item[5)]{$H_1$ and $H_2$ have constant row and column wights,}
\item[6)]{For each choice of $R$ satisfying Criteria 3) and 4), $R$ has the constant column weight.}
\end{itemize}
We call a CSS-type code obtained by a pair of binary matrices meeting all the six criteria an \textit{adaptive quantum noise control code} (AQNCC).
If a channel always produces asymmetric noise, one might wish to exchange rows in the same manner by using $H_1$ and $H_2$
defining asymmetric quantum LDPC codes in which no choice of $R$ gives exactly the same error correction abilities for the two types of errors.
We call such CSS-type codes \textit{askew} AQNCCs.
The parameters of AQNCCs will be referred to in the same manner as EAQECCs.

Now we present a combinatorial construction for AQNCCs.
A \textit{cyclic difference matrix} (CDM) of \textit{order} $v$ with $\mu$ rows, denoted by $(v,\mu)$ CDM, is a $\mu \times v$ matrix $M = (m_{i,j})$
with entries from the cyclic group $\mathbb{Z}_v$ of order $v$
such that for each $0 \leq i < j \leq \mu-1$, the set $\{m_{i,\ell} - m_{j,\ell} | 0 \leq \ell \leq v-1\}$ contains every element of $\mathbb{Z}_v$.
We assume that the elements of $\mathbb{Z}_v$ are represented by nonnegative integers up to $v-1$ by taking the residue group of order $v$.
\begin{theorem}\label{CDM}
For any odd prime $p \geq 5$ and any integer $i$, $0 \leq i \leq \frac{p-5}{2}$, there exists a $[[p^2,2(i+1)(p-1);1]]$ \textup{AQNCC}.
\end{theorem}
\Proof
Consider the set of $p-$tuple vectors $\{\boldsymbol{r}_a\}$, where $\boldsymbol{r}_a = (0, a, 2a, \dots, (p-1)a)$, $1 \leq a \leq p-1$, over the ring $\mathbb{Z}_p$.
The $(p-1) \times p$ matrix obtained by stacking $\boldsymbol{r}_a$ forms a $(p,p-1)$ CDM.
Let $I(x)$ be the circulant permutation matrix with a one at the $(x+y)$th column and the $y$th row.
For each $x \in \mathbb{Z}_v$, replace all entries $x$ in the CDM with $I(x)$.
Let $H_1$ be the first $\frac{p-1}{2}$ layers of $p$ circulant permutation matrices and $H_2$ the remaining $\frac{p-1}{2}$ layers.
Two different rows of which one is from $H_1$ and the other from $H_2$ share a one at exactly one position.
The row weight is uniformly $p$. Thus, we have $\rk{H_1H_2^T} = 1$.
Because two different rows within $H_1$ or $H_2$ share a one at most one position, $H_1$ and $H_2$ have no $4$-cycles.
It is easy to check that $\boldsymbol{r}_{p-a} = (0, (p-1)a, (p-2)a, \dots, a)$.
Hence $H_1$ and $H_2$ are parity-check matrices of regular LDPC codes with girth six which are nonidentical and isomorphic.
Take an arbitrary set $R$ of $r$ layers from $H_2$ for some nonnegative integer $r \leq \frac{p-1}{2}-1$
and name the resulting $(\frac{p(p-1)}{2}-pr) \times p^2$ matrix as $H_2'$.
Add $R$ to $H_1$ to create a $(\frac{p(p-1)}{2}+pr) \times p^2$ matrix $H_1'$.
It is straightforward to see that $\rk{H_1'H_2'^T} = 1$ regardless of the choice of $R$.
The row and column weights of $R$ are constant.
Thus, $H_1'$ and $H_2'$ give entanglement-assisted quantum regular LDPC codes requiring only one ebit with girth six again.
A simple linear algebraic calculation proves that the rank of any $j$ layers of $p$ circulant permutation matrices in the expanded CDM is $j(p-1)+1$.
Hence, we have $\rk{H_1} + \rk{H_2} = \rk{H_1'} + \rk{H_2'}$. Thus, $H_1$ and $H_2$ give a $[[p^2,2(p-1);1]]$ AQNCC.
Discarding $i$ layers each from $H_1$ and $H_2$ increases the dimension by $2i(p-1)$.
\qed

\begin{theorem}\label{askewCDM}
For any odd prime $p \geq 3$ and any integer $i$, $0 \leq i \leq \frac{p-3}{2}$, there exists an askew $[[p^2,(2i+3)(p-1);1]]$ \textup{AQNCC}.
\end{theorem}
\Proof
Construct a $(p,p-1)$ CDM as in the proof of Theorem~\ref{CDM} and put $\boldsymbol{r}_0 = (0, 0, \dots, 0)$ on top of the matrix.
The resulting matrix is a $(p,p)$ CDM. Expanding the CDM and choosing $R$ as in the proof of Theorem \ref{CDM} prove the assertion.
\qed

Here we give a small example of AQNCCs given in Theorem \ref{CDM}.
We first construct a $(7,6)$ CDM using rows $\boldsymbol{r}_i = (0, i, 2i, 3i, 4i, 5i, 6i)$ over $\mathbb{Z}_7$ as follows:
\[\left(\begin{array}{c}
\boldsymbol{r}_1\\
\boldsymbol{r}_2\\
\boldsymbol{r}_3\\
\boldsymbol{r}_4\\
\boldsymbol{r}_5\\
\boldsymbol{r}_6\\
\boldsymbol{r}_7
\end{array}\right)
=
\left(\begin{array}{ccccccc}
0&1&2&3&4&5&6\\
0&2&4&6&1&3&5\\
0&3&6&2&5&1&4\\
0&4&1&5&2&6&3\\
0&5&3&1&6&4&2\\
0&6&5&4&3&2&1
\end{array}\right).\]
Replace entry $x$ with $I(x)$ to create a $42 \times 49$ binary matrix. For example, every ``$2$" in the above matrix is replaced by
\[\left(\begin{array}{ccccccc}
0&0&1&0&0&0&0\\
0&0&0&1&0&0&0\\
0&0&0&0&1&0&0\\
0&0&0&0&0&1&0\\
0&0&0&0&0&0&1\\
1&0&0&0&0&0&0\\
0&1&0&0&0&0&0
\end{array}\right).\]
Let $H_1$ be the first three layers of $7$ circulant permutation matrices obtained by replacing the entries of the first three rows of the CDM.
$H_2$ is obtained by replacing the entries of the remaining half of the CDM. Applying $H_1$ and $H_2$ to the CSS construction gives
entanglement-assisted quantum LDPC codes with girth six of parameters $[[49,12;1]]$.
We can turn this code into asymmetric codes without changing the length and dimension. For example, moving the layer coming from $\boldsymbol{r}_4$ or the two layers coming from $\boldsymbol{r}_4$ and $\boldsymbol{r}_5$ to $H_1$, we obtain parameters $[[49,12;1]]$ or $[[49,12;1]]$ respectively.
If the noise level of a channel becomes lower, one can increase the dimension of the AQNCC by deleting layers from $H_1$ or $H_2$ or both.



\section{Performance}\label{sec:simulation}

\begin{figure}
  \includegraphics[width=0.5\textwidth]{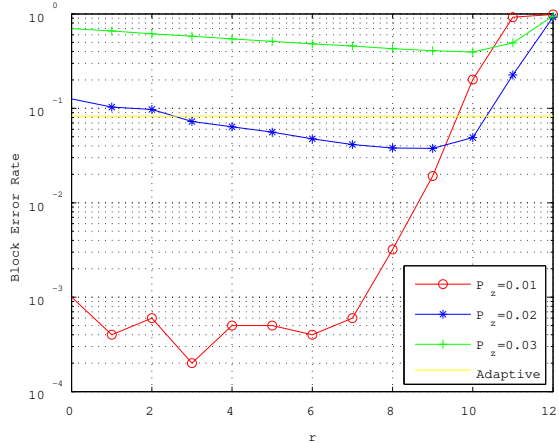}\\
  \caption{Block error performance of the $[[841,56;1]]$ AQNCC constructed from Theorem~\ref{CDM} where we choose $p=29$ and $i=0$. The vertical axis represents the block error rate. The horizontal axis represents the extra correcting power of the phase error over the bit error. Each $r$ in the horizontal axis corresponds to a $[[841,56;1]]$ AQNCC where the matrix $H_1'$ contains extra $r$ layers coming from $H_2$ in the CSS construction in Theorem~\ref{CDM}. The AQNCC corresponds to a standard EAQECC (same correcting power between the phase errors and the bit errors) when $r=0$. The phase errors $P_z$ in the asymmetric Pauli channel range from $0.01$ to $0.03$ while the bit error $P_x$ is fixed at $0.005$. In the adaptive simulation, the phase error $P_z$ is slowly time-variant while the bit error $P_x$ is time-invariant. Specifically, we assume that $P_z$ changes every $100$ uses of the channels, and is uniformly distributed between $0$ and $0.03$. }\label{error_plot}
\end{figure}

We performed a series of numerical simulations of the AQNCC constructed in Sec.~\ref{sec:adaptive noise control}. In particular, we chose a medium size $[[841,56;1]]$ AQNCC constructed according to Theorem~\ref{CDM} (with $p=29$), and depict the block error performance in Fig.~\ref{error_plot}. In the simulations, we used the iterative decoding algorithm since our AQNCCs are also sparse quantum codes. The correcting power of the phase errors over the bit errors in our AQNCCs is controlled by the difference of ranks of the parity check matrices $H_1'$ and $H_2'$ (see the proof of Theorem~\ref{CDM} for the definition of $H_1'$ and $H_2'$) . When $\text{rank}(H_1')=\text{rank}(H_2')$, our AQNCC corresponds to the standard QECC, and are suitable for the symmetric channels.
As one might expect from carefully designed asymmetric QECCs,
a notable result is that the AQNCCs perform much better than the standard EAQECCs when the Pauli channel is asymmetric. For example, the best block error rate among the set of AQNCCs is four times better than the standard EAQECC when $P_z=0.02$ and $P_x=0.005$.

We can adjust our AQECCs by changing the value of $\text{rank}(H_1')-\text{rank}(H_2')$ while keeping the block size and the pre-shared entanglement the same. This allows us to use our AQNCCs in an adaptive manner. We consider the simplest case where the phase error $P_z$ is slowly time-variant, but the bit error $P_x$ is time-invariant. Specifically, we assume that $P_z$ changes every $100$ uses of the channels, and is uniformly distributed between $0$ and $0.03$. The receiver will notify the sender to increase the phase correcting power in the AQNCCs if he fails to decode the phase errors but can correctly decode the bit errors.

\section{Conclusion}\label{sec:conclusion}
We have described principles of a flexible error-correction protocol, adaptive quantum noise control,
which can optimize error correction performance of quantum error-correcting codes according to the characteristics of channels during information transmission.
The primary theoretical tools are the entanglement-assisted stabilizer formalism, low-density parity-check codes, and combinatorial design theory.
The combination of the three allowed us to design quantum error-correcting codes which can flexibly trade error correction abilities for phase errors and bit errors.
Our method requires only one ebit, which would make it easier to implement adaptive noise control in the future.
Because physical devices are expected to cause phase errors much more frequently,
we believe that adaptive quantum noise control code will be of importance in various situations.

An interesting question is whether a similar flexible optimization can be realized without using ebits.
While it seems to be difficult without ebits because of the severe limitation of the symplectic orthogonality, it is certainly worth investigation.
Designing adaptive quantum noise control codes for more parameters with better overall performance is also an interesting open problem.
We expect that the extensive use of information theory and combinatorial design theory to analyze parity-check matrices as in \cite{FCVBT,FC,CF}
will be key to a further development of the theory of adaptive noise control.


\end{document}